\documentclass[aps,pra,twocolumn,amsmath,amssymb,superscriptaddress]{revtex4}

\newcommand{\bra}[1]{\langle#1|}
\newcommand{\ket}[1]{|#1\rangle}

\usepackage[dvips]{graphicx}
\usepackage{mathrsfs}

\begin{document}

\bibliographystyle{apsrev}

\title{Comparison of architectures for approximating number-resolving photo-detection using non-number-resolving detectors}

\author{Peter P. Rohde}
\email[]{rohde@physics.uq.edu.au}
\homepage{http://www.physics.uq.edu.au/people/rohde/}
\affiliation{These authors made an equal contribution to this paper}
\affiliation{Centre for Quantum Computer Technology, Department of Physics\\ University of Queensland, Brisbane, QLD 4072, Australia}

\author{James G. Webb}
\affiliation{These authors made an equal contribution to this paper}
\affiliation{Centre for Quantum Computer Technology, School of Information Technology and Electrical Engineering, University College, The University of New South Wales, Canberra ACT 2600, Australia}

\author{Elanor H. Huntington}
\affiliation{Centre for Quantum Computer Technology, School of Information Technology and Electrical Engineering, University College, The University of New South Wales, Canberra ACT 2600, Australia}

\author{Timothy C. Ralph}
\affiliation{Centre for Quantum Computer Technology, Department of Physics\\ University of Queensland, Brisbane, QLD 4072, Australia}

\date{\today}

\frenchspacing

\begin{abstract}
Number-resolving photo-detection is necessary for many quantum optics experiments, especially in the application of entangled state preparation. Several schemes have been proposed for approximating number-resolving photo-detection using non-number-resolving detectors. Such techniques include multi-port detection and time-division multiplexing. We provide a detailed analysis and comparison of different number-resolving detection schemes, with a view to creating a useful reference for experimentalists. We show that the ideal architecture for projective measurements is a function of the detector's dark count and efficiency parameters. We also describe a process for selecting an appropriate topology given actual experimental component parameters.
\end{abstract}

\pacs{42.50.-p}

\maketitle

\section{Introduction}

Number-resolving photo-detection is a necessary prerequisite for many important quantum optics applications, most notably in the emerging field of optical quantum information processing \cite{bib:Kok05, bib:Ralph06}. Unfortunately, presently available photo-detectors are incapable of resolving photon number with high fidelity. In fact, most commonly available photo-detectors are so-called `bucket' or `on/off' detectors, which can distinguish only between two cases -- no photons, and one or more photons. These practical limitations in the number-resolving capabilities of photo-detectors have motivated the development of techniques for approximating number-resolving detection using non-number-resolving detectors. Most notably, such techniques include multi-port networks \cite{bib:Kok01, bib:Paul96, bib:Bartlett02, bib:Rohde05}, time-division multiplexing (TDM) \cite{bib:Achilles03, bib:Achilles04, bib:Banaszek03, bib:Fitch03, bib:Pereira07} and visible light photon counting modules (VLPC's) \cite{bib:Kim99, bib:Takeuchi99, bib:Bartlett02}. All of these techniques are variations on a single simple idea -- the optical field is distributed across multiple modes which are measured independently. For a large number of modes the probability of any given mode being populated by more than one photon approaches zero. Thus, the sum of the detection events across the modes closely approximates the number of photons in the incident state.

In this paper we analyze and compare different architectures for implementing photon-number-resolving detection. Our analysis is primarily concerned with the performance of photon-number-resolving projective measurements.  Such measurements are central to the preparation of, for example, one mode of a bipartite entangled state by the detection of the other as described in \cite{bib:Rubin00}.  Whilst photon number resolving detectors are also useful for the reconstruction of photostatistics, this application is not considered in this paper.  Statistical reconstruction including the effects of the most dominant practical effects -- loss and dark-counts have been extensively studied in \cite{bib:Zambra05} and \cite{bib:Lee04}.

We begin by considering a very general setting, a lossy multi-port detector cascade \cite{bib:Kok01}. Most architectures for implementing number-resolving detection we are aware of reduce to specific instances of this setting. First we derive an expression for the conditional probability $P(m|n)$ of detecting $m$ photons given an incident $n$-photon Fock state.  Our analysis includes the experimental effects of loss, finite detection efficiency and dark counts neglected in other similar treatments \cite{bib:Achilles04}. We omit consideration of after-pulsing, as this may be disregarded with appropriate detector timing. We then derive a positive operator value measure (POVM) description of the measurement process, which is directly related to these conditional probabilities. The POVM description can be related to a quantum process description, which provides a very general characterization that can be applied to understand the dynamics of many different systems.

Using this approach we characterize and contrast the $N$-port detector, and two primary variants of time division multiplexed (TDM) detectors -- the balanced TDM detector \cite{bib:Fitch03} and the loop detector \cite{bib:Banaszek03} from a projective measurement perspective. Finally, we conclude with a design procedure for selecting an appropriate architecture and then optimising the design. We demonstrate that given contemporary experimental component parameters, the balanced TDM architecture is generally most useful.

\section{General analysis}

\begin{figure}[!htb]
\includegraphics[width=0.4\columnwidth]{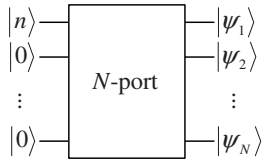}
\caption{$N$-port detector cascade. A state is incident upon one of the inputs, while vacuum states are input into the remaining inputs. The $N$-port then applies some transformation which distributes the incident field across the outputs $1\hdots N$. Finally, the outputs are detected independently using non-number-resolving detectors.} \label{fig:N_port}
\end{figure}

We begin by considering a very general setting as follows. We have an $N$-port detector cascade with an $n$-photon Fock state incident upon one of the inputs (see Fig.~\ref{fig:N_port}). The input couples to the outputs with coupling efficiency $p_c(i)$ between the input and output $i$. That is, a given photon has probability $p_c(i)$ of being directed to output $i$. The paths through the interferometer are lossy. The probability of a photon following the path to output $i$ being lost is $p_\mathrm{loss}(i)$. Following the $N$-port, each output is measured independently using a bucket detector \footnote{Our treatment of photo-detectors ignores spatio-temporal effects such as finite bandwidth and dead-time \cite{bib:RohdeRalph06b}. Instead we restrict ourselves to their behavior in the photon-number degree of freedom.}.  Each detector is subject to a dark-count probability of $p_\mathrm{dc}$. We do not need to explicitly introduce a detector efficiency term, since this can be absorbed into the $p_\mathrm{loss}$ terms.

\subsection{Measurement probability analysis}

Upon detection of the outputs of the cascade we will observe a particular pattern of detector \emph{clicks}. We refer to a particular combination of clicks as a \emph{signature}. Thus, a signature is simply a binary sequence of length $N$. Let us consider the probability of a particular signature occurring. We use the notation $P_\mathrm{sig}(\vec{d})$ to denote the probability of the detection signature occurring whereby each of the detectors in vector $\vec{d}$ trigger, and all others do not. This probability is given by \footnote{MATLAB implementations of these expressions may be freely downloaded from our website: http://seal.tst.adfa.edu.au/research/quantelec/matlab/}
\begin{eqnarray}
P_\mathrm{sig}(\vec{d}) &=& \sum_{n_1+\dots+n_N=n} \frac{n!}{n_1! \dots n_N!} \, \prod_{i=1}^N p_c(i)^{n_i} \nonumber\\
&\times& \prod_{i\in \vec{d}} \left[p_\mathrm{dc} + (1 - p_\mathrm{dc}) [1 - p_\mathrm{loss}(i)^{n_i}] \right] \nonumber\\
&\times& \prod_{i\notin \vec{d}} \left[(1 - p_\mathrm{dc}) p_\mathrm{loss}(i)^{n_i} \right].
\end{eqnarray}
The first line of this expression is a sum over all possible configurations of how the $n$ photons can reach the $N$ outputs (before loss), weighted by the probability of each event. The variables $n_i$ are the number of photons that reach the $i^\mathrm{th}$ output, before loss. The second and third lines represent the probabilities of the desired signature occurring, given the respective configuration.

The net probability of detecting $m$ photons is given by summing over all combinations of detection signatures where $|\vec{d}|=m$, where $|\vec{d}|$ is total number of clicks. Thus,
\begin{equation}
P(m|n) = \sum_{|\vec{d}|=m}P_\mathrm{sig}(\vec{d}).
\end{equation}
Note that a complete set of probabilities $P(m|n)\,\forall\,m,n$ completely characterizes the operation of the detector.

It is worth noting that $P_\mathrm{sig}(\vec{d})$ is not necessarily the same for all $\vec{d}$ where $|\vec{d}|=m$. That is, different signatures corresponding to the same measured number of photons needn't have equal probabilities of occurring. This will be the case, for example, in loop-based time-division-multiplexers, where there is an inherent asymmetry in photon arrival probabilities across different time-bins. This is something we will discuss in detail in later sections.

Qualitatively one can make several observations about any scheme which is an instance of this general scenario. First, in the ideal case ($p_\mathrm{loss}=p_\mathrm{dc}=0$) we expect the $N$-port to approach an ideal detector in the limit \mbox{$N\to\infty$} and $p_c(i)\to 0\,\,\forall\,\,i$. In other words, in the limit where the incident field is split into an infinite number of infinitely small components the probability of multiple photons reaching a single detector approaches zero and the probability of correctly measuring photon number approaches unity. For $p_\mathrm{dc}>0$ however this will no longer be the case, since each output port introduces an extra opportunity for a dark-count to occur. Thus, for large $N$ false counts become a certainty. Therefore, for any given application one expects there to be an optimal value of $N$ subsequently referred to as $N_\mathrm{opt}$, which depends upon the experimental parameters. Based on this observation one intuitively expects that for small $p_\mathrm{dc}$ it will be better to use large $N$, whereas for large $p_\mathrm{dc}$ it will be better to use smaller $N$.

\subsection{POVM \& quantum process descriptions}

We have derived a general expression for the operation of a lossy $N$-port detector in terms of the conditional probability $P(m|n)$. For many applications a simple expression for the conditional probability is insufficient. Most notably, when using a detector to implement projective photon number measurements, a simple probability measure is insufficient to derive the form of the projected state. For this a description of the measurement process in terms of measurement operators is necessary. We now consider the POVM description of the measurement process implemented by this general detector. We denote the POVM element corresponding to the $m$-photon detection outcome by $\hat\Pi(m)$. This POVM element takes the form
\begin{equation}
\hat\Pi(m) = \sum_n P(m|n) \ket{n}\bra{n}.
\end{equation}
Also, we implicitly assume that the measured state is traced out after measurement.  This is because the measurement process destroys the incident state. Note that for an ideal detector (i.e. $N\to\infty$, $p_c\to 0$, $p_\mathrm{loss}=p_\mathrm{dc}=0$), we have $P(m|n)=\delta_{m,n}$, in which case the POVM elements reduce to
\begin{equation}
\hat\Pi_\mathrm{ideal}(m) = \ket{m}\bra{m},
\end{equation}
as expected for an ideal number-resolving detector. In the general case however, these POVM's result in mixing over different possible measurement outcomes. This occurs for three independent reasons: loss means that an $n$-photon state may be confused for a \mbox{$<n$}-photon state; dark-counts mean that an $n$-photon state may be confused for a \mbox{$>n$}-photon state; and finite $N$ means that the probability of more than one photon appearing at a given output is non-zero, again meaning that an $n$-photon state may be confused for a \mbox{$<n$}-photon state.

The measurement process may also be described as a quantum process. When expressed in the basis of photon number projectors these processes are characterized by diagonal process matrices,
\begin{eqnarray}
\mathcal{E}_m(\hat\rho) &=& \sum_{ij} \chi_{ij}^{(m)} \hat{E}_i\hat\rho\hat{E}_j\nonumber\\
&=& \sum_{n} P(m|n) \hat{E}_n\hat\rho\hat{E}_n
\label{eq:process_matrix}
\end{eqnarray}
where $\chi$ is a process matrix and $\hat{E}_n=\ket{n}\bra{n}$ are the photon number projectors.  In the ideal case where $P(m|n)=\delta_{m,n}$, the process matrix corresponding to this quantum process will be the zero matrix with a single `1' at the $m^\mathrm{th}$ location along the main diagonal.  Thus, $[\chi(m)]_{i,j} = \delta_{i,j}\delta_{j,m}$ and Eq.~\ref{eq:process_matrix} reduces to
\begin{eqnarray}
\mathcal{E}_m(\hat\rho) &=& \hat{E}_m\hat\rho\hat{E}_m
\end{eqnarray}

\section{Example detection architectures}

In this section we apply our general analysis to several specific well-known architectures that have been experimentally demonstrated. We first consider the case of a balanced $N$-port detector, perhaps the best known scheme for implementing number-resolving photo-detection. Then we consider two variations on time-division-multiplexed photo-detection. These two variations are distinct in that their loss and coupling characteristics are inherently different. In one case the coupling terms are uniform, while in the other they are necessarily non-uniform.  The relevant parameters for the different detection architectures considered are thus summarized in Table~\ref{table:summary}.
\begin{table}[!htb]
\begin{tabular}{|c|c|c|}
\hline
Architecture      & $p_c(i)$                                    & $p_\mathrm{loss}(i)$          \\
\hline
Balanced $N$-port & $1/N$                                                & $p_\mathrm{loss}$             \\
Loop TDM          & $p_c(1-p_c)^{i-1}$                 & $1-t_s^i t_c^i t_f^{i-1}\eta_\mathrm{det}$ \\
Balanced TDM      & $1/N$                                                & $1-t_f^{i-1}t_c^{m+1}\eta_\mathrm{det}$     \\
\hline
\end{tabular}
\caption{Summary of relevant parameters for different detection architectures. The parameters $t_\mathrm{f}$, $t_\mathrm{c}$ and $t_\mathrm{s}$ indicate the transmission (i.e. $1-$loss) of the optical fibre, coupler and switch components illustrated in Figs.~\ref{fig:loop_TDM} and \ref{fig:balanced_TDM} respectively.  $m$ is the number of stages of the balanced TDM detector.  $\eta_\mathrm{det}$ is the quantum efficiency of the final non photon number resolving detector used.}
\label{table:summary}
\end{table}

\subsection{Balanced $N$-port detection} \label{sec:balanced_N_port_detection}

We first consider the case of a balanced $N$-port detector of Fig.~\ref{fig:N_port}. Here an incident state is distributed equally across $N$ outputs and thus the coupling parameters are all equal, $p_c(i) = 1/N \,\,\forall\,\, i$.  If it is assumed that the device is internally constructed such that loss is uniform across the outputs (i.e. by using a tree network of ideal beamsplitters), $p_\mathrm{loss}(i) = p_\mathrm{loss} \,\,\forall\,\, i$.  Such a symmetric device has been previously analysed in detail in \cite{bib:Kok01}.

In practise, due to their complexity, experimentalists do not construct $N$-ports directly. Instead they build setups which closely approximate $N$-port detection. Perhaps the closest approximation is the visible light photon counter (VLPC). Here an incident light field is spread out spatially and incident upon are large detector consisting of many small active detection areas which may trigger independently. Although this provides a close approximation of a balanced $N$-port, it is imperfect due to asymmetry in the distribution of the field across the detector regions. Specifically, the light field incident upon the detector will have a roughly 2D Gaussian distribution, rather than a uniform distribution. Nonetheless, in the regime where the number of detector regions is large compared to the number of incident photons this provides a good approximation of number-resolving photo-detection.

\subsection{Loop time-division multiplexed detection} \label{sec:loop_TDM_detection}

\begin{figure}[!htb]
\includegraphics[width=0.6\columnwidth]{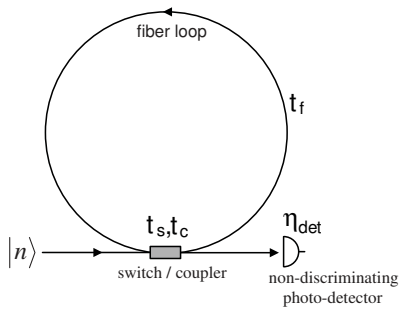}
\caption{Loop time-division multiplexing. A state enters the fiber on the left and is coupled into the fiber loop by the electro-optic switch / coupler. It then repeatedly circulates through the fiber loop. Following each round-trip photons have some probability of coupling out to the photo-detector through the coupler. Thus, the state is divided into discrete time-bins, which are independently detected.  $t_\mathrm{f}$ refers to the optical transmission of the fibre loop.  $t_\mathrm{c}$ is the transmission of the coupler, including connector losses.  $t_\mathrm{s}$ indicates the transmission through the switch.} \label{fig:loop_TDM}
\end{figure}

Now we consider the loop-detector illustrated in Fig.~\ref{fig:loop_TDM}. This architecture has previously been considered in Ref. \cite{bib:Pereira07}. Here the optical state is coupled into a fibre loop. The loop couples out again via a coupler with coupling strength $p_c$. Thus, after each round-trip of the loop photons have a probability $p_c$ of coupling out to the photo-detector.

We relate the loop-detector to the generalized $N$-port interferometer by letting the $i^\mathrm{th}$ port of the interferometer represent the corresponding time-bin of the loop-detector.  Since photons in the $i^\mathrm{th}$ (where $i \ge 1$) time-bin must undergo $i-1$ round-trips and make $i$ transits of the switch and coupler we have \mbox{$p_\mathrm{loss}(i) = 1-t_s^i t_c^i t_f^{i-1}\eta_\mathrm{det}$} and \mbox{$p_c(i) = p_c (1-p_c)^{i-1}$}.

In principle a loop detector can continue detecting for an arbitrarily long time span, thereby effectively implementing $N\to\infty$. In practise however, because of the exponential decay in the coupling term with the number of round-trips, it is sufficient to truncate measurement and consider a relatively small number of time-bins. In fact, in the presence of dark-count it is preferable to truncate the number of measured time-bins. The reason for this is that the contribution of legitimate counts drops exponentially against the number of round-trips, whereas the dark-count rate stays constant. Thus, for large $N$ the later time-bins will achieve nothing other than to contribute unwanted dark-counts.

Truncation gives rise to errors. Specifically, there will be some probability that photons land in the truncated region, which are discarded by the analysis, giving rise to erroneous results. In an experimental context this makes the measurement results ambiguous, reducing the fidelity of the measured state, while in a theoretical context this introduces an error margin in the analysis. Thus, one must be careful when considering results for the loop TDM to ensure that we are in a regime where the probability of photons landing in the truncated region is small. Classically, this probability is of the order
\begin{equation} \label{eq:error_margin}
P_\mathrm{error} = (1-p_c)^N.
\end{equation}
For quantum optical states this probability may be much lower, depending upon the likelihood of photon numbers $>N$.

\subsection{Balanced time-division multiplexed detection} \label{sec:balanced_TDM_detection}

\begin{figure}[!htb]
\includegraphics[width=\columnwidth]{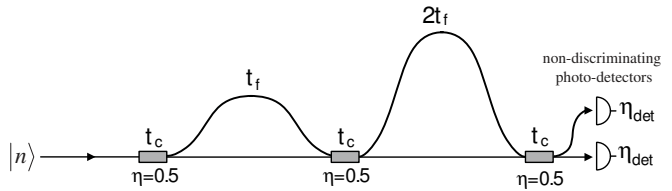}
\caption{A balanced TDM setup for $m=2$ stages (i.e. $N=4$). Thus, photons are distributed equally across four possible paths.  $t_\mathrm{f}$ and $t_\mathrm{c}$ are as defined previously, in Fig.~\ref{fig:loop_TDM}.} \label{fig:balanced_TDM}
\end{figure}

The final architecture we consider is that of balanced time-division multiplexing, shown in Fig.~\ref{fig:balanced_TDM}. This architecture consists of $m$ stages (not to be confused with $m$, the number of measured photons -- the context will make the distinction clear), giving rise to $N=2^m$ distinct non-overlapping time bins and is topologically equivalent to a balanced $N$-port. That is, the coupling to each time bin is equal and $p_c(i) = 1/N\,\,\forall\,\,i$. Despite being topologically equivalent, the balanced TDM differs from balanced $N$-port detection in that loss rates are not uniform across each time bin. Specifically, photons reaching the $i^\mathrm{th}$ bin will pass through $i-1$ lengths of fiber and $m+1$ 50:50 couplers.  Hence we have \mbox{$p_\mathrm{loss}(i) = 1-t_f^{i-1}t_c^{m+1}\eta_\mathrm{det}$}.

While the $N$-port and loop TDM architectures exhibit an optimal number of bins $N_\mathrm{opt}$, which optimizes fidelity, and a minimum number of bins required to resolve the desired photon number, $N_\mathrm{min}$, the balanced TDM architecture exhibits an optimal and minimum number of stages $m_\mathrm{opt}$ and $m_\mathrm{min}$ respectively.

\section{Analysis of architectures for the conditional preparation of Fock states via parametric down-conversion} \label{sec:parametric_down_conv}

To introduce our experimental example and to expand on the general outline provided in the previous section, we now consider a common scenario; conditional preparation of specific Fock states via non-degenerate parametric down-conversion. This example is very important to present-day experiments where this is the defacto standard for the preparation of single photons. Here we condition on detecting some number of photons in one arm of the down-converter. When conditioning succeeds we expect an equal number of photons to be present in the other output due to the photon number correlations between the two output modes.

The output state of a non-degenerate down-converter takes the form \cite{bib:GerryKnight05}
\begin{equation}
\ket\psi = \frac{1}{\sqrt{1-\chi^2}}\sum_n \chi^n \ket{n}\ket{n},
\end{equation}
where $\chi$ is a parameter related to the down-conversion strength (not to be confused with the process matrix we introduced previously). Next we apply the $m$ photon measurement process, $\mathcal{E}_m$, to obtain the conditioned state,
\begin{eqnarray}
\hat\rho_\mathrm{cond} &=& \mathcal{E}_m(\ket\psi\bra\psi)\nonumber\\
&=& \frac{1}{1-\chi^2} \sum_n P(m|n) \chi^{2n} \ket{n}\bra{n}.
\end{eqnarray}
Following renormalization, the prepared state is given by
\begin{equation}
\hat\rho_\mathrm{prep} = \frac{\hat\rho_\mathrm{cond}}{\mathrm{tr}(\hat\rho_\mathrm{cond})} = \frac{\sum_n P(m|n) \chi^{2n} \ket{n}\bra{n}}{\sum_n P(m|n) \chi^{2n}}.
\end{equation}
To quantify how well the state preparation procedure works, we calculate the fidelity between the the conditionally prepared state and the expected state $\ket{m}$,
\begin{equation} \label{eq:fidelity}
F(\hat\rho_\mathrm{cond},\ket{m}) = \frac{P(m|m) \chi^{2m}}{\sum_n P(m|n) \chi^{2n}}.
\end{equation}
In the limit of ideal photo-detection, $P(m|n)=\delta_{m,n}$, the fidelity reduces to unity, as expected. Notice that that non-zero $P(m|m)$, as is the case for non-ideal photo-detection, adversely affects the detection fidelity. The above fidelity only provide a measure of how sure we are of the desired state upon a detection event. The actual probability of the desired $\bra{m}$ projection event $p_\mathrm{det}$ is given by
\begin{equation} \label{eq:probability}
p_\mathrm{det} = \sum_n P(m|n)|\langle n|\psi\rangle|^2 = \frac{1}{{1-\chi^2}} \sum_n {P(m|n)\chi^{2n}}.
\end{equation}
In the case of non-ideal detection, $P(m|m)$ will in general be $<1$ and the terms $P(m|n)$ ($m \neq n$) will be non-zero.

\subsection{Balanced $N$-port detection}

First let us consider the operation of this setup using a balanced $N$-port configuration for the conditioning detector.

\begin{figure}[!htb]
\includegraphics[width=\columnwidth]{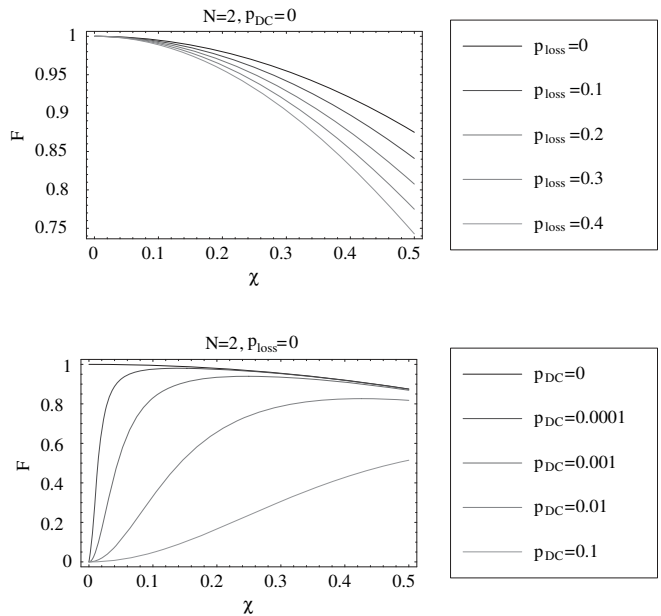}
\caption{Fidelity against loss (top) and dark-count (bottom) rates for fixed $N=2$.} \label{fig:N_port_eta}
\end{figure}

In Fig. \ref{fig:N_port_eta} we plot the fidelity of projection onto a single photon state against $\chi$ for different values of $p_\mathrm{loss}$, where we only assume detector loss, and $p_\mathrm{dc}$. As expected, the fidelity decreases monotonically with both these parameters.
In the absence of dark-counts the fidelity drops monotonically with $\chi$. This is because as $\chi$ increases so too does the probability of higher photon number terms arising. In the presence of dark-counts, we observe maxima in $F$ against $\chi$. The initial increase in $F$ is because dark-count rates are constant whereas the probability of a single photon number term occurring is monotonically increasing. Thus, when the probability of a single photon term occurring is very small the effects of dark-counts will dominate, reducing fidelity. As $\chi$ increases so too does the effect of the desired single photon terms. Then for even higher values of $\chi$ the probability of higher photon number terms becomes significant, also reducing the fidelity.

\begin{figure}[!htb]
\includegraphics[width=\columnwidth]{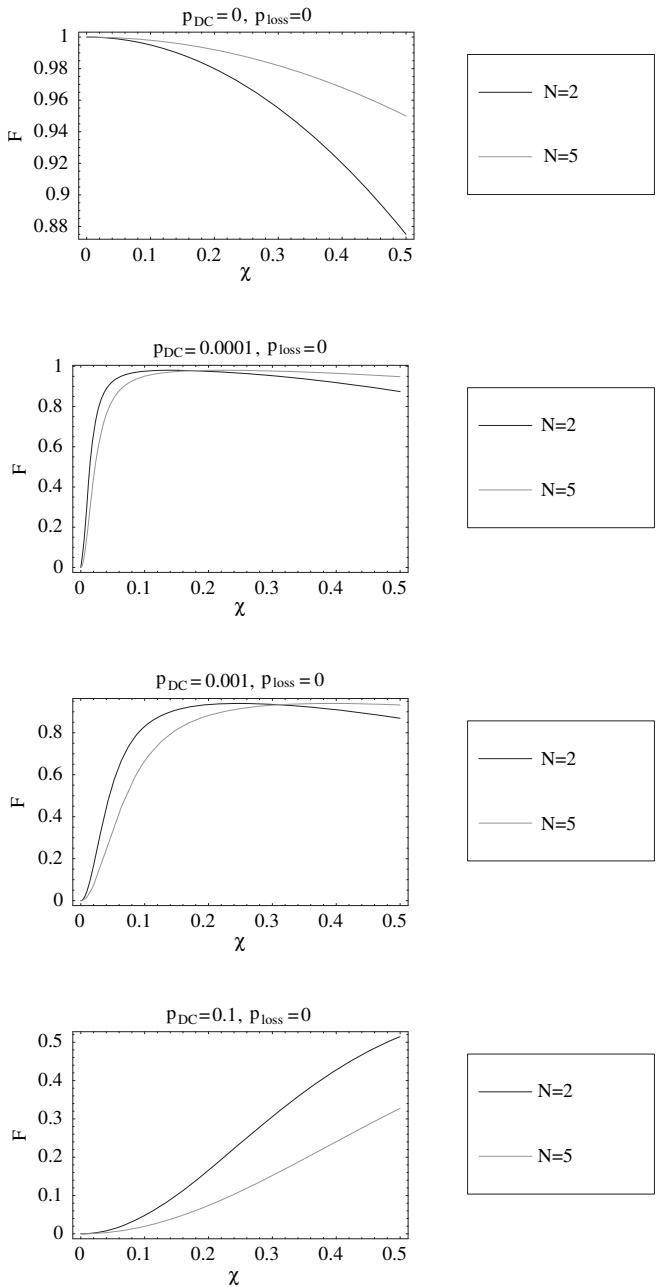}
\caption{Comparison of the fidelity of a conditionally prepared state via non-degenerate parametric down-conversion for the balanced $N$-port detection architecture.} \label{fig:plots_N_port}
\end{figure}

In Fig. \ref{fig:plots_N_port} we plot the fidelity of the conditionally prepared state against down-conversion strength, $\chi$, for $N=2$ and $N=5$, and various values of $p_\mathrm{dc}$.
In the ideal case, $p_\mathrm{dc}=p_\mathrm{loss}=0$, it is beneficial to use larger $N$ since this increases the confidence when discriminating between different number states. However, in the presence of dark-counts it is not strictly beneficial to use larger $N$, since increasing $N$ also increases the probability of a dark-count occurring.

\begin{figure}[!htb]
\includegraphics[width=0.7\columnwidth]{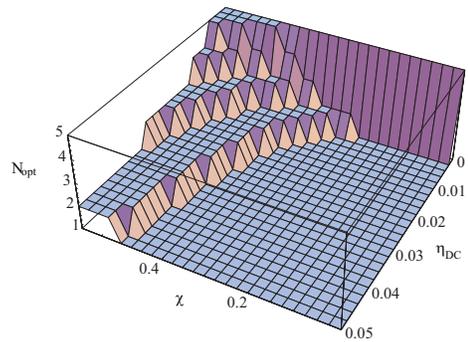}
\caption{Optimal value of $N$ against down-conversion strength $\chi$, and dark-count rate $p_\mathrm{dc}$ for a balanced $N$-port configuration. $p_\mathrm{loss}=0$.} \label{fig:optimal_N_balanced}
\end{figure}

This observation motivates us to determine $N_\mathrm{opt}$ for given loss and dark-count parameters. In Fig. \ref{fig:optimal_N_balanced} we plot the value of $N_\mathrm{opt}$ (i.e. the value of $N$ that optimized $F$) against $\chi$ and $p_\mathrm{dc}$. There are two important trends taking place. First, as the down-conversion strength increases, it becomes increasingly desirable to increase $N$. This is because as $\chi$ increases so too does the probability of generating higher photon numbers, thus making the additional number resolving power of higher $N$ necessary. Second, as the dark-count rate increases it becomes desirable to use smaller $N$. This is because the additional ports increase the probability of a dark-count occurring. We have not included a plot against $p_\mathrm{loss}$ in this case, since it's effect on $N_\mathrm{opt}$ is close to uniform and almost negligible for the range of $\chi$ and $p_\mathrm{dc}$ considered.

\subsection{Loop time-division multiplexing}

Next we consider the performance of loop TDM detection in the context of conditional photon preparation. The inherent asymmetry in the loop TDM architecture suggests there may also be asymmetry in the fidelity of different detection signatures corresponding to the same number of photons. For example, consider the case where there is no loss, but some dark-counts occurring. The dark-count probability is uniform across the detectors, but the probability distribution of incident photons is biased towards the earlier time-bins. Thus, a detection event occurring in an earlier time-bin will have higher relative probability of being caused by an incident photon than if the detection had occurred at the second time-bin.

\begin{figure}[!htb]
\includegraphics[width=0.8\columnwidth]{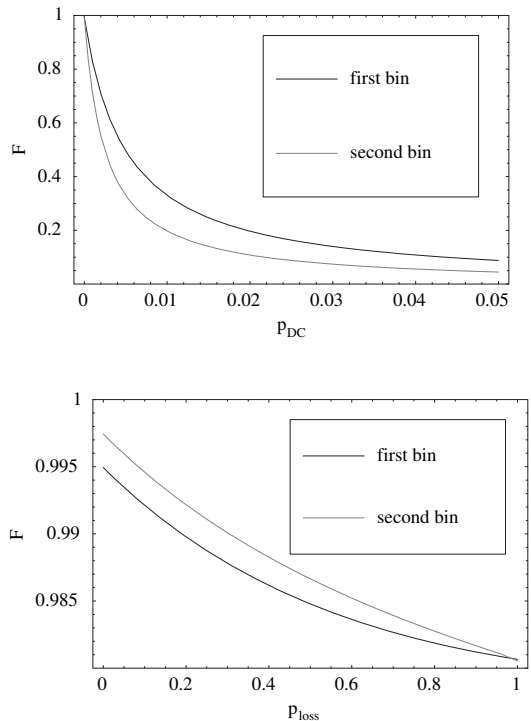}
\caption{Asymmetry in the performance of different detection events corresponding to $|\vec{d}|=1$ in the loop TDM architecture. $\chi=0.1$, $p_c=0.5$.} \label{fig:assymetric_sigs}
\end{figure}

To illustrate this, in Fig. \ref{fig:assymetric_sigs} we plot the fidelity against $p_\mathrm{dc}$ and $t_\mathrm{f}$ (and assume $t_\mathrm{c}, t_\mathrm{s}= 1$) for the $\{1,0,0,0,0\}$ and $\{0,1,0,0,0\}$ signatures, both of which correspond to detection of a single photon.
In the presence of dark-counts we indeed observe that it is favorable to condition on the $\{1,0,\dots\}$ signature than $\{0,1,\dots\}$. In the presence of loss the converse is true. However, this does not arise because of the loss itself. Rather it is inherent -- higher order time bins have exponentially lower coupling probabilities, and therefore stronger number resolving power.

Thus, loss and dark-counts are competing parameters when determining the optimal conditioning signature. However, note that the dependence of fidelity on dark-counts is much greater than on loss. There are two reasons for this. First, there is a multiplicative effect from dark-counts -- i.e. every detector is independently subject to dark-counts. Thus, in our simulation where $N=5$, there will be a roughly 5 times multiplicative factor in the dependence on dark-counts. This highlights the necessity of truncating loop TDM to minimize dark-count multiplication. Second, loss causes $n>1$ terms to be confused as $n=1$ terms. However, because $n>1$ terms have very low probability, the fidelity does not suffer very much from these confusions. This is in contrast to the dark-count effect, which occur independently of how probable the $n>1$ terms are.

\subsection{Balanced time-division-multiplexing}
Finally we consider operation when using balanced TDM detection. Qualitatively the behavior is very similar to the $N$-port considered previously. In particular, in the presence of dark-counts there are distinct regions where different values of $N$ are optimal. In the presence of no fiber loss (but potentially detector loss), we do not observe that the fidelity varies significantly across different signatures corresponding to the same number of detected photons (the graphics for this have not been included). This arises because this system is almost balanced, so ideally the confidence is distributed evenly across the bins. There is a slight deviation from this because, although coupling rates are uniformly distributed, loss rates are not. However, in this application this only affects the confidence between distinguishing one photon from higher photon number. This occurs with very low probability, so this effect is not noticeable.

\section{Example design procedure} \label{sec:example_procedure}

We now consider the details of experimental implementation of the projective measurements from the previous section. We omit the $N$-port for reasons of impracticality \footnote{An $N$-port setup requires $N$ independent photo-detectors, which becomes completely impractical for large $N$.} and describe the optimal design of both a balanced TDM system and a loop architecture detector at both 780 nm and 1550 nm wavelengths.  The assumed experimental parameters for the two wavelengths are shown in Table~\ref{table:parameters}, using values typical of contemporary single mode fibre optic components. The detectors are assumed to be operating with 20ns gate windows and with dead times less than the fibre delay $\tau$.

\begin{table}[!htb]
\begin{tabular}{|c|c|c|}
\hline
Component & 780 nm & 1550 nm  \\
\hline
Coupler loss   & 0.4 dB               & 0.5 dB \\
Fibre loss     & 0.2 dB               & 0.8 dB \\
               & $L=10$m ($\tau=50$ns)& $L=2$km ($\tau=10\mu$s)\\
Switch loss    & 2.0 dB               & 1.2 dB \\
\hline
               & Silicon              & InGaAs \\
               & Perkin-Elmer         & id-Quantique \\
 Detector      & SPCM-AQR-13-FC       &   id200\\
               & $\eta_\mathrm{det} = 60\%$        & $\eta_\mathrm{det} = 10\%$         \\
               & $p_\mathrm{dc}=5\times10^{-6}$ & $p_\mathrm{dc}=9.6\times10^{-4}$\\
\hline

\end{tabular}
\caption{Assumed experimental parameters for the optical components at 780 nm and 1550 nm.}
\label{table:parameters}
\end{table}

\begin{figure}[!htb]
\includegraphics[width=\columnwidth]{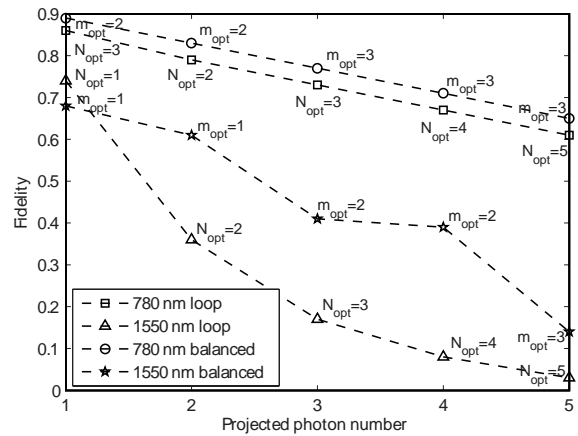}
\caption{Maximum achievable fidelity for the two TDM architectures for projection onto the Fock state shown. $N_\mathrm{opt}=N_\mathrm{min}$ for all loop detector data points (except $\langle 1|$, 780 nm loop). For the loop detectors the number of bins is truncated to $N_\mathrm{opt}$. $m_\mathrm{opt}$ = optimum number of stages for balanced TDM architecture. For the loop TDM's the coupling ratios were chosen according to Eq. \ref{eq:error_margin} so as to bound truncation error to 1\%. They were 0.99, 0.90, 0.78, 0.68 and 0.60 for $N=1\dots 5$ respectively.}
\label{fig:f_vs_n}
\end{figure}

To perform a projection onto an $n$ photon Fock state, we require $N \ge n$. Thus irrespective of architecture, the minimum number of detection bins is $N_\mathrm{min} = n$, with $N_\mathrm{opt} \ge N_\mathrm{min}$.  The loop detector exhibits the additional free parameter $p_c$ which may be optimised for a given application. Fig.~\ref{fig:f_vs_n} illustrates the optimum projection fidelity achieved by both topologies at both wavelengths with $\chi=0.3$. In the case of the two loop TDM lines we employ the coupling ratio that saturates the error bound from Eq. \ref{eq:error_margin}, where we impose a maximum error rate of 1\%. The reason for doing this is as follows. For a loop TDM we always wish to minimize $p_c$, since this distributes the incident field across the largest number of bins. However, as $p_c$ is lowered the probability of photons being truncated increases. Thus, we minimize $p_c$ subject to the constraint that some error bound be satisfied. In the case of the balanced TDM, we perform a search over $m$, the number of stages, such that the fidelity is maximized. The experimental parameter values found to optimise the projected fidelity are listed beside each data point.

\begin{figure}[!htb]
\includegraphics[width=\columnwidth]{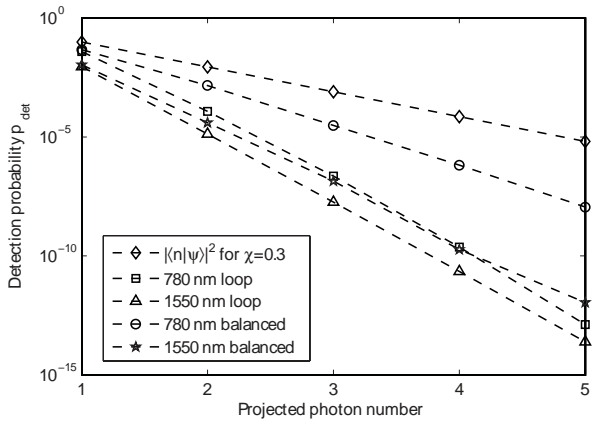}
\caption{Probability of projecting onto desired Fock state. $|\langle n|\psi\rangle|^2$ represents the probability of finding $n$ photons in the parametric downconverter output state. Note the logarithmic axes. For loop detectors the number of bins is truncated to $N_{opt}$. The upper line illustrates the performance of an ideal number resolving detector. This represents an upper bound on what is achievable and puts the other lines into perspective.}
\label{fig:det_prob}
\end{figure}

Fig.~\ref{fig:det_prob} illustrates the corresponding $p_\mathrm{det}$ for each of the detection topologies and projections shown in Fig.~\ref{fig:f_vs_n}.

\subsection{Loop TDM detector}

Apparent from the above example is the surprising empirical result that generally $N_\mathrm{opt} = N_\mathrm{min}$. This is a consequence of dark-counts. As $N$ increases the overall probability of being affected by a dark-count increases. Thus it is desirable to keep $N$ as small as possible, provided it is sufficiently large to measure $n$, i.e. $N_\mathrm{min}$.
From Fig. \ref{fig:f_vs_n} it is evident that the projection fidelity decreases with photon number. The plot of $|\langle n|\psi\rangle|^2$ in Fig.~\ref{fig:det_prob} illustrates the low probability with which these terms occur, however. Fig.~\ref{fig:det_prob} also reveals a vanishingly small $p_\mathrm{det}$ for both loop detectors considered.

Because the measurement is truncated to $N$ time bins, there is some probability of photons not reaching the detector. Thus it would thus appear desirable to increase $N$ to improve the overall detection probability. This, however requires a reduction of the total loop loss (i.e. $t_f t_c t_s \to 1$) to counter the effects of dark counts in order to maintain the projection fidelity.

Consequently, there exist certain combinations of detector and loop parameters whereby optimum projection fidelities are obtained when $N_\mathrm{opt} > N_\mathrm{min}$.  It is difficult to describe these general conditions due to their dependance upon both $\ket{\psi}$ and $\bra{n}$.  For example, with $\chi=0.3$ for the parametric down converted state considered, $N_\mathrm{opt}=3$ for the 780 nm loop detector projection onto $\bra{1}$ (as shown in Fig.~\ref{fig:f_vs_n}).  With $\chi=0.15$ it is observed that $N_\mathrm{opt} = N_\mathrm{min}$ for all projections.  Simulated results have typically indicated that fidelities within 1\% of the maximum attainable for all values of $N$ are achieved with $N_\mathrm{opt} = N_\mathrm{min}$.

\subsection{Balanced TDM detector}

As $N=2^m$, the minimum number of stages $m_\mathrm{min}$ required to perform a projection onto $\ket{n}$ is given by $m_\mathrm{min} \ge \lceil\log_2{n}\rceil$.  For the 1550 nm balanced data plotted in Figs.~\ref{fig:f_vs_n} and~\ref{fig:det_prob} $m_\mathrm{opt}=m_\mathrm{min}$.  The consequence of $N$ only being able to assume powers of two is evident in the fidelities of the projections onto $\ket{1}$, $\ket{3}$ and $\ket{5}$ appearing lower than the general trend would suggest.  Notice that the projections performed by the silicon detector benefit from $m_\mathrm{opt} > m_\mathrm{min}$. This is because this detector has a sufficiently low dark-count rate that larger values of $N$ do not become corrupted by the increasing dark-count probability.

\begin{figure}[!htb]
\includegraphics[width=\columnwidth]{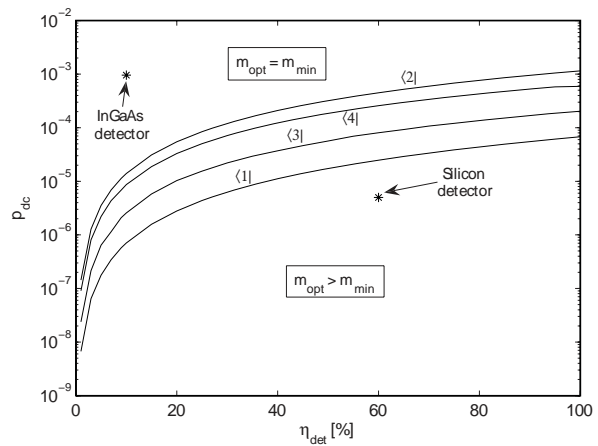}
\caption{Plot defining the detector parameters required to benefit from $m_\mathrm{opt} > m_\mathrm{min}$ ($N_\mathrm{opt} > N_\mathrm{min}$) for projection onto the Fock states shown, using a balanced TDM architecture.  $\chi = 0.3$ and the balanced detector component losses are as defined for 780 nm in Table~\ref{table:parameters}. The characteristic detector parameters of $p_\mathrm{dc}$ and $\eta_\mathrm{det}$ are illustrated for our example detectors.  Note the logarithmic axes.}
\label{fig:bal_proj}
\end{figure}

Fig.~\ref{fig:bal_proj} indicates the combinations of $p_\mathrm{dc}$ and $\eta_\mathrm{det}$ which favour $m_\mathrm{opt} > m_\mathrm{min}$ for the projective measurements indicated.  Note that projection onto power of two photon numbers places the most relaxed constraints on $p_\mathrm{dc}$.  This is a result of satisfying $N_\mathrm{min} = n$ and thus the effects of dark counts being minimized. It is clearly seen that the silicon detector benefits from $m_\mathrm{opt} > m_\mathrm{min}$ in our application for all values of $n$, but the InGaAs detector does not.  It should be noted that at the time of writing there are other InGaAs detectors available such as \footnote{id-Quantique offer an 'Ultra-Low noise' variant of the id201. $p_\mathrm{dc}$ may also typically be reduced by reducing the gate duration.} which offer better values of $p_\mathrm{dc}$ and $\eta_\mathrm{det}$ than our example. $m_\mathrm{opt} > m_\mathrm{min}$ may be a worthwhile option with such detectors.

\subsection{Design considerations}

Experimentalists are generally faced with the problem of designing a system to meet a desired specification, given constraints imposed by the components available.  An example design scenario might be to determine the detector parameters $\eta_\mathrm{det}$, $p_\mathrm{dc}$ required to achieve a given projection fidelity.  Given the finite number of detector parameter combinations available, however, a more likely problem is how to maximise the fidelity for practical values of $\eta_\mathrm{det}$, $p_\mathrm{dc}$ and component losses.

While the performance of the two architectures scale differently with respect to $\eta_\mathrm{det}$ and $p_\mathrm{dc}$, two general observations may be made.  Firstly, minimisation of $p_\mathrm{dc}$ is of key importance to maximise fidelity, particularly when projecting onto higher photon number states.  Secondly, both schemes degrade relatively slowly with decreasing $\eta_\mathrm{det}$.  Consequently in the selection of a detector, preference should be given to its dark noise performance.

To select an appropriate topology, the calculations of Eqs.~\ref{eq:fidelity} and~\ref{eq:probability} should be performed for both systems with component losses applicable to the wavelength of operation and $N=N_\mathrm{min}$ initially.  To determine the optimum performance attainable from the loop architecture, firstly calculate the value of $p_c$ corresponding to the maximum desired truncation error.  Secondly, iterate $p_c$ through all available greater experimental coupling ratios, noting the value corresponding to greatest projection fidelity.  The above process should be repeated for $N > N_\mathrm{min}$ when particularly low values of $p_\mathrm{dc}$ and/or when significant probabilities of higher photon number terms exist.  A choice may thus be made between the topologies on the grounds of projection fidelity and $p_\mathrm{det}$.

Overall, given these criteria, the best choice for a detector to perform projective measurements is likely to be the balanced TDM architecture.  However, both schemes offer useful projection fidelities and careful consideration of $\chi$, $\eta_\mathrm{det}$, $p_\mathrm{dc}$ and component losses is required to select and optimise the performance of a given architecture.  Either architecture is suitable for the purposes of reconstruction of photon statistics, with the loop detector having a potential advantage given that $N$ may be increased arbitrarily to limits imposed by the reconstruction process and $p_\mathrm{dc}$. In general, the presence of loss, finite detection efficiencies and dark counts only influence the number of independent measurements required to arbitrarily minimize statistical errors \cite{bib:Lee04}.

\section{Conclusion}

In our analysis we have studied the three dominant architectures for implementing photon number resolving photo-detection using non-number resolving detectors. We first considered the archetypal protocol, balanced $N$-port detection. For this architecture we considered the effects of loss, dark-counts and the number of bins $N$. We also considered how to optimize this architecture under different experimental conditions. We then turned out attention to two forms of time-division multiplexing -- the loop TDM, and balanced TDM. We compared these two schemes and how to optimize them. We also included results using experimentally realistic parameters.

Our analysis has considered the major experimental limitations of the balanced $N$-port detector cascade, and the loop and balanced TDM detectors. While we have investigated the performance of the three architectures from the perspective of projective measurements made on one mode of an optical parametric down converter, our treatment is sufficiently general that our conclusions may be extended to any projective scenario.

The balanced TDM system appears most resilient with respect to non-ideal detector quantum efficiency and dark noise performance at the cost of requiring an additional non photon number resolving detector. The loop detector remains a useful architecture, however only for certain combinations of experimental parameters and projection operations.

\begin{acknowledgments}
This work was supported by the Australian Research Council and Queensland State Government. We acknowledge partial support by the DTO-funded U.S. Army Research Office Contract No. W911NF-05-0397.
\end{acknowledgments}

\bibliography{paper}

\begin{thebibliography}{18}
\expandafter\ifx\csname natexlab\endcsname\relax\def\natexlab#1{#1}\fi
\expandafter\ifx\csname bibnamefont\endcsname\relax
  \def\bibnamefont#1{#1}\fi
\expandafter\ifx\csname bibfnamefont\endcsname\relax
  \def\bibfnamefont#1{#1}\fi
\expandafter\ifx\csname citenamefont\endcsname\relax
  \def\citenamefont#1{#1}\fi
\expandafter\ifx\csname url\endcsname\relax
  \def\url#1{\texttt{#1}}\fi
\expandafter\ifx\csname urlprefix\endcsname\relax\def\urlprefix{URL }\fi
\providecommand{\bibinfo}[2]{#2}
\providecommand{\eprint}[2][]{\url{#2}}

\bibitem[{\citenamefont{Kok et~al.}(2005)\citenamefont{Kok, Munro, Ralph,
  Dowling, and Milburn}}]{bib:Kok05}
\bibinfo{author}{\bibfnamefont{P.}~\bibnamefont{Kok}},
  \bibinfo{author}{\bibfnamefont{W.~J.} \bibnamefont{Munro}},
  \bibinfo{author}{\bibfnamefont{T.~C.} \bibnamefont{Ralph}},
  \bibinfo{author}{\bibfnamefont{J.~P.} \bibnamefont{Dowling}},
  \bibnamefont{and} \bibinfo{author}{\bibfnamefont{G.~J.}
  \bibnamefont{Milburn}} (\bibinfo{year}{2005}), \eprint{quant-ph/0512071}.

\bibitem[{\citenamefont{Ralph}(2006)}]{bib:Ralph06}
\bibinfo{author}{\bibfnamefont{T.~C.} \bibnamefont{Ralph}},
  \bibinfo{journal}{Rep. Prog. Phys.} \textbf{\bibinfo{volume}{69}},
  \bibinfo{pages}{853} (\bibinfo{year}{2006}).

\bibitem[{\citenamefont{Kok and Braunstein}(2001)}]{bib:Kok01}
\bibinfo{author}{\bibfnamefont{P.}~\bibnamefont{Kok}} \bibnamefont{and}
  \bibinfo{author}{\bibfnamefont{S.~L.} \bibnamefont{Braunstein}},
  \bibinfo{journal}{Phys. Rev. A} \textbf{\bibinfo{volume}{63}},
  \bibinfo{pages}{033812} (\bibinfo{year}{2001}).

\bibitem[{\citenamefont{Paul et~al.}(1996)\citenamefont{Paul, T{\"o}rm{\"a},
  Kiss, and Jex}}]{bib:Paul96}
\bibinfo{author}{\bibfnamefont{H.}~\bibnamefont{Paul}},
  \bibinfo{author}{\bibfnamefont{P.}~\bibnamefont{T{\"o}rm{\"a}}},
  \bibinfo{author}{\bibfnamefont{T.}~\bibnamefont{Kiss}}, \bibnamefont{and}
  \bibinfo{author}{\bibfnamefont{I.}~\bibnamefont{Jex}},
  \bibinfo{journal}{Phys. Rev. Lett.} \textbf{\bibinfo{volume}{76}},
  \bibinfo{pages}{2464} (\bibinfo{year}{1996}).

\bibitem[{\citenamefont{Bartlett et~al.}(2002)\citenamefont{Bartlett, Diamanti,
  Sanders, and Yamamoto}}]{bib:Bartlett02}
\bibinfo{author}{\bibfnamefont{S.~D.} \bibnamefont{Bartlett}},
  \bibinfo{author}{\bibfnamefont{E.}~\bibnamefont{Diamanti}},
  \bibinfo{author}{\bibfnamefont{B.~C.} \bibnamefont{Sanders}},
  \bibnamefont{and} \bibinfo{author}{\bibfnamefont{Y.}~\bibnamefont{Yamamoto}},
  \bibinfo{journal}{Proceedings of Free-Space Laser Communication and Laser
  Imaging II} \textbf{\bibinfo{volume}{4821}} (\bibinfo{year}{2002}).

\bibitem[{\citenamefont{Rohde}(2005)}]{bib:Rohde05}
\bibinfo{author}{\bibfnamefont{P.~P.} \bibnamefont{Rohde}},
  \bibinfo{journal}{J. Opt. B} \textbf{\bibinfo{volume}{7}},
  \bibinfo{pages}{82} (\bibinfo{year}{2005}).

\bibitem[{\citenamefont{Achilles et~al.}(2003)\citenamefont{Achilles,
  Silberhorn, \'Sliwa, Banaszek, and Walmsley}}]{bib:Achilles03}
\bibinfo{author}{\bibfnamefont{D.}~\bibnamefont{Achilles}},
  \bibinfo{author}{\bibfnamefont{C.}~\bibnamefont{Silberhorn}},
  \bibinfo{author}{\bibfnamefont{C.}~\bibnamefont{\'Sliwa}},
  \bibinfo{author}{\bibfnamefont{K.}~\bibnamefont{Banaszek}}, \bibnamefont{and}
  \bibinfo{author}{\bibfnamefont{I.~A.} \bibnamefont{Walmsley}},
  \bibinfo{journal}{Opt. Lett.} \textbf{\bibinfo{volume}{28}},
  \bibinfo{pages}{2387} (\bibinfo{year}{2003}).

\bibitem[{\citenamefont{Achilles et~al.}(2004)\citenamefont{Achilles,
  Silberhorn, Sliwa, Banaszek, Walmsley, Fitch, Jacobs, Pittman, and
  Franson}}]{bib:Achilles04}
\bibinfo{author}{\bibfnamefont{D.}~\bibnamefont{Achilles}},
  \bibinfo{author}{\bibfnamefont{C.}~\bibnamefont{Silberhorn}},
  \bibinfo{author}{\bibfnamefont{C.}~\bibnamefont{Sliwa}},
  \bibinfo{author}{\bibfnamefont{K.}~\bibnamefont{Banaszek}},
  \bibinfo{author}{\bibfnamefont{I.~A.} \bibnamefont{Walmsley}},
  \bibinfo{author}{\bibfnamefont{M.~J.} \bibnamefont{Fitch}},
  \bibinfo{author}{\bibfnamefont{B.~C.} \bibnamefont{Jacobs}},
  \bibinfo{author}{\bibfnamefont{T.~B.} \bibnamefont{Pittman}},
  \bibnamefont{and} \bibinfo{author}{\bibfnamefont{J.~D.}
  \bibnamefont{Franson}}, \bibinfo{journal}{J. Mod. Opt.}
  \textbf{\bibinfo{volume}{51}}, \bibinfo{pages}{1499} (\bibinfo{year}{2004}).

\bibitem[{\citenamefont{Banaszek and Walmsley}(2003)}]{bib:Banaszek03}
\bibinfo{author}{\bibfnamefont{K.}~\bibnamefont{Banaszek}} \bibnamefont{and}
  \bibinfo{author}{\bibfnamefont{I.}~\bibnamefont{Walmsley}},
  \bibinfo{journal}{Opt. Lett.} \textbf{\bibinfo{volume}{28}},
  \bibinfo{pages}{52} (\bibinfo{year}{2003}).

\bibitem[{\citenamefont{Fitch et~al.}(2003)\citenamefont{Fitch, Jacobs,
  Pittman, and Franson}}]{bib:Fitch03}
\bibinfo{author}{\bibfnamefont{M.~J.} \bibnamefont{Fitch}},
  \bibinfo{author}{\bibfnamefont{B.~C.} \bibnamefont{Jacobs}},
  \bibinfo{author}{\bibfnamefont{T.~B.} \bibnamefont{Pittman}},
  \bibnamefont{and} \bibinfo{author}{\bibfnamefont{J.~D.}
  \bibnamefont{Franson}}, \bibinfo{journal}{Phys. Rev. A}
  \textbf{\bibinfo{volume}{68}}, \bibinfo{pages}{043814}
  (\bibinfo{year}{2003}).

\bibitem[{\citenamefont{The and Ramos}(2007)}]{bib:Pereira07}
\bibinfo{author}{\bibfnamefont{G.~A.~P.} \bibnamefont{The}} \bibnamefont{and}
  \bibinfo{author}{\bibfnamefont{T.~V.} \bibnamefont{Ramos}},
  \bibinfo{journal}{J. Mod. Opt.} \textbf{\bibinfo{volume}{54}},
  \bibinfo{pages}{1187} (\bibinfo{year}{2007}).

\bibitem[{\citenamefont{Kim et~al.}(1999)\citenamefont{Kim, Takeuchi, Yamamoto,
  and Hogue}}]{bib:Kim99}
\bibinfo{author}{\bibfnamefont{J.}~\bibnamefont{Kim}},
  \bibinfo{author}{\bibfnamefont{S.}~\bibnamefont{Takeuchi}},
  \bibinfo{author}{\bibfnamefont{Y.}~\bibnamefont{Yamamoto}}, \bibnamefont{and}
  \bibinfo{author}{\bibfnamefont{H.~H.} \bibnamefont{Hogue}},
  \bibinfo{journal}{App. Phys. Lett.} \textbf{\bibinfo{volume}{74}},
  \bibinfo{pages}{902} (\bibinfo{year}{1999}).

\bibitem[{\citenamefont{Takeuchi et~al.}(1999)\citenamefont{Takeuchi, Kim,
  Yamamoto, and Hogue}}]{bib:Takeuchi99}
\bibinfo{author}{\bibfnamefont{S.}~\bibnamefont{Takeuchi}},
  \bibinfo{author}{\bibfnamefont{J.}~\bibnamefont{Kim}},
  \bibinfo{author}{\bibfnamefont{Y.}~\bibnamefont{Yamamoto}}, \bibnamefont{and}
  \bibinfo{author}{\bibfnamefont{H.~H.} \bibnamefont{Hogue}},
  \bibinfo{journal}{App. Phys. Lett.} \textbf{\bibinfo{volume}{74}},
  \bibinfo{pages}{1063} (\bibinfo{year}{1999}).

\bibitem[{\citenamefont{Rubin}(2000)}]{bib:Rubin00}
\bibinfo{author}{\bibfnamefont{M.~H.} \bibnamefont{Rubin}},
  \bibinfo{journal}{Phys. Rev. A} \textbf{\bibinfo{volume}{61}},
  \bibinfo{pages}{022311} (\bibinfo{year}{2000}).

\bibitem[{\citenamefont{Zambra et~al.}(2005)\citenamefont{Zambra, Andreoni,
  Bondani, Gramegna, Genovese, Brida, Rossi, and Paris}}]{bib:Zambra05}
\bibinfo{author}{\bibfnamefont{G.}~\bibnamefont{Zambra}},
  \bibinfo{author}{\bibfnamefont{A.}~\bibnamefont{Andreoni}},
  \bibinfo{author}{\bibfnamefont{M.}~\bibnamefont{Bondani}},
  \bibinfo{author}{\bibfnamefont{M.}~\bibnamefont{Gramegna}},
  \bibinfo{author}{\bibfnamefont{M.}~\bibnamefont{Genovese}},
  \bibinfo{author}{\bibfnamefont{G.}~\bibnamefont{Brida}},
  \bibinfo{author}{\bibfnamefont{A.}~\bibnamefont{Rossi}}, \bibnamefont{and}
  \bibinfo{author}{\bibfnamefont{M.~G.~A.} \bibnamefont{Paris}},
  \bibinfo{journal}{Phys. Rev. Lett.} \textbf{\bibinfo{volume}{95}},
  \bibinfo{pages}{063602} (\bibinfo{year}{2005}).

\bibitem[{\citenamefont{Lee et~al.}(2004)\citenamefont{Lee, Yurtsever, Kok,
  Hockney, Adami, Braunstein, and Dowling}}]{bib:Lee04}
\bibinfo{author}{\bibfnamefont{H.}~\bibnamefont{Lee}},
  \bibinfo{author}{\bibfnamefont{U.}~\bibnamefont{Yurtsever}},
  \bibinfo{author}{\bibfnamefont{P.}~\bibnamefont{Kok}},
  \bibinfo{author}{\bibfnamefont{G.~M.} \bibnamefont{Hockney}},
  \bibinfo{author}{\bibfnamefont{C.}~\bibnamefont{Adami}},
  \bibinfo{author}{\bibfnamefont{S.~L.} \bibnamefont{Braunstein}},
  \bibnamefont{and} \bibinfo{author}{\bibfnamefont{J.~P.}
  \bibnamefont{Dowling}}, \bibinfo{journal}{J. Mod. Opt.}
  \textbf{\bibinfo{volume}{51}}, \bibinfo{pages}{1517} (\bibinfo{year}{2004}).

\bibitem[{\citenamefont{Gerry and Knight}(2005)}]{bib:GerryKnight05}
\bibinfo{author}{\bibfnamefont{C.~C.} \bibnamefont{Gerry}} \bibnamefont{and}
  \bibinfo{author}{\bibfnamefont{P.~L.} \bibnamefont{Knight}},
  \emph{\bibinfo{title}{Introductory quantum optics}}
  (\bibinfo{publisher}{Cambridge University Press}, \bibinfo{year}{2005}).

\bibitem[{\citenamefont{Rohde and Ralph}(2006)}]{bib:RohdeRalph06b}
\bibinfo{author}{\bibfnamefont{P.~P.} \bibnamefont{Rohde}} \bibnamefont{and}
  \bibinfo{author}{\bibfnamefont{T.~C.} \bibnamefont{Ralph}},
  \bibinfo{journal}{J. Mod. Opt.} \textbf{\bibinfo{volume}{53}},
  \bibinfo{pages}{1589} (\bibinfo{year}{2006}).

\end{thebibliography}

\end{document}